\documentclass[10pt,a4paper]{article}
\usepackage[margin=1in]{geometry}
\usepackage{graphicx}
\usepackage{amsmath,amssymb}
\usepackage{cite}
\usepackage{authblk}
\usepackage{hyperref}

\title{\textbf{Constraints on Higgs Light Yukawa Couplings with the CMS Detector}}
\author[]{Alberto Zucchetta\textsuperscript{1}, on behalf of the CMS Collaboration}
\affil[]{\textsuperscript{1}INFN Sezione di Padova, Via Marzolo 8, Padova, Italy}

\date{Presented at the 32\textsuperscript{nd} International Symposium on Lepton Photon Interactions at High Energies, Madison, Wisconsin, USA, August 25-29, 2025}

\begin{document}
\maketitle

\begin{abstract}
The discovery of the Higgs boson ten years ago and successful measurement of the Higgs boson couplings to third generation fermions by LHC mark great milestones for HEP. The much weaker coupling to the second generation quarks predicted by the SM makes the measurement of the light Yukawa Higgs couplings, as the Higgs-charm ones, much more challenging. With the latest tagging algorithms capabilities, a lot of progress has been made to constrain these couplings. In this talk, the latest results of direct and indirect measurements by the CMS experiment are presented. Prospects for future improvements are also given.
\end{abstract}

\section{Introduction}
The observation of the Higgs boson at the LHC in 2012~\cite{discoveryATLAS,discoveryCMS,discoverylongCMS} confirmed the existence of the particle responsible for electroweak symmetry breaking. Since then, one of the main goals of the ATLAS and CMS Collaborations has been to accurately measure its couplings to gauge bosons and fermions. While couplings to third-generation fermions (top and bottom quarks, and the tau lepton) and gauge bosons have been firmly established, the Yukawa interactions involving lighter quarks remain largely unconstrained due to their small expected branching ratios and challenging experimental signatures.

In the SM, the Yukawa coupling of a fermion $f$ is proportional to its mass, $y_f = \sqrt{2} m_f/v$, where $v=246$~GeV is the Higgs vacuum expectation value. Many extensions of the SM predict deviations from this relation, thus measuring or constraining the charm and light-quark Yukawa couplings provides an essential test of the SM.
In order to quantify the deviation from the SM expected values, a multiplicative factor $k$ is introduced to parametrize deviations from the Higgs boson couplings, such that $y_{f,V} = k_{f,V} \cdot y_{f,V}^{SM}$. In the SM, all $k$ factors correspond to unity.

The CMS experiment~\cite{cms} explores these couplings through several complementary approaches:\ (1) direct searches for $H\to c\bar{c}$ decays,\ (2) searches for Higgs production in association with a single charm quark ($cH$),\ and (3) indirect measurements of the well established $H\to ZZ\to 4\ell$ decay, and the search for rare decays such as $H\to J/\Psi\,\gamma$, which proceeds through quantum loops involving charm quarks.

\section{Direct Searches for $H\to c\bar{c}$}
The most direct approach to constrain $y_c$ is through the measurement of the Higgs decay to charm quark pairs. Due to the overwhelming QCD multijet background, this search is only feasible in production channels with additional event-level discrimination, such as associated production with a vector boson ($VH$) or top quarks ($t\bar{t}H$).

\subsection{Higgs Production in Association with Top Quarks}
The $t\bar{t}H$ production mode provides a challenging probe of $H\to c\bar{c}$ decays, because the final state is characterized by multiple jets (up to eight), leptons (up to two), and neutrinos, and includes $b$- and $c$-jets. Because of the intrinsic resolution of the hadronic jets, resonances ($t$, $W$, $H$) cannot be unambiguously reconstructed.

The CMS analysis~\cite{HIG24018}, based on Run-2 data, considers all possible final states and develops a novel approach based on the state-of-the-art multi-class jet flavor tagging algorithm and a sophisticated neural network to classify the events.

The jet flavor tagging algorithm, known as \texttt{ParticleNet}~\cite{ParticleNet2020}, is a customized neural network architecture using Dynamic Graph Convolutional Neural Network that assigns probabilities for each jet to originate from a $b$-, $c$-, or light-flavor quark based on low-level information derived from the unordered jet constituents.
This algorithm represents a significant improvement, up to a factor 2, over the previous-generation tagger (known as \texttt{DeepJet}) in terms of separation between light- and heavy flavor jets, and between $b$- and $c$-jets.

The CMS analysis considers 9 event classes, consisting of 4 different signals ($t\bar{t}H$, $t\bar{t}Z$ in the decays $H\to b\bar{b}$ and  $H\to c\bar{c}$) and 5 background classes, which include the production of a top quark pair in association with single or multiple additional light- or heavy-flavor quarks ($b$ and $c$).

A transformer-based neural network combines the event-level kinematic variables to estimate the probability of the event to belong to each class. The network outputs are then used to separate events into categories, and to serve as the main discriminant in the statistical inference.

\begin{figure}[ht]
\centering
\includegraphics[width=0.62\textwidth]{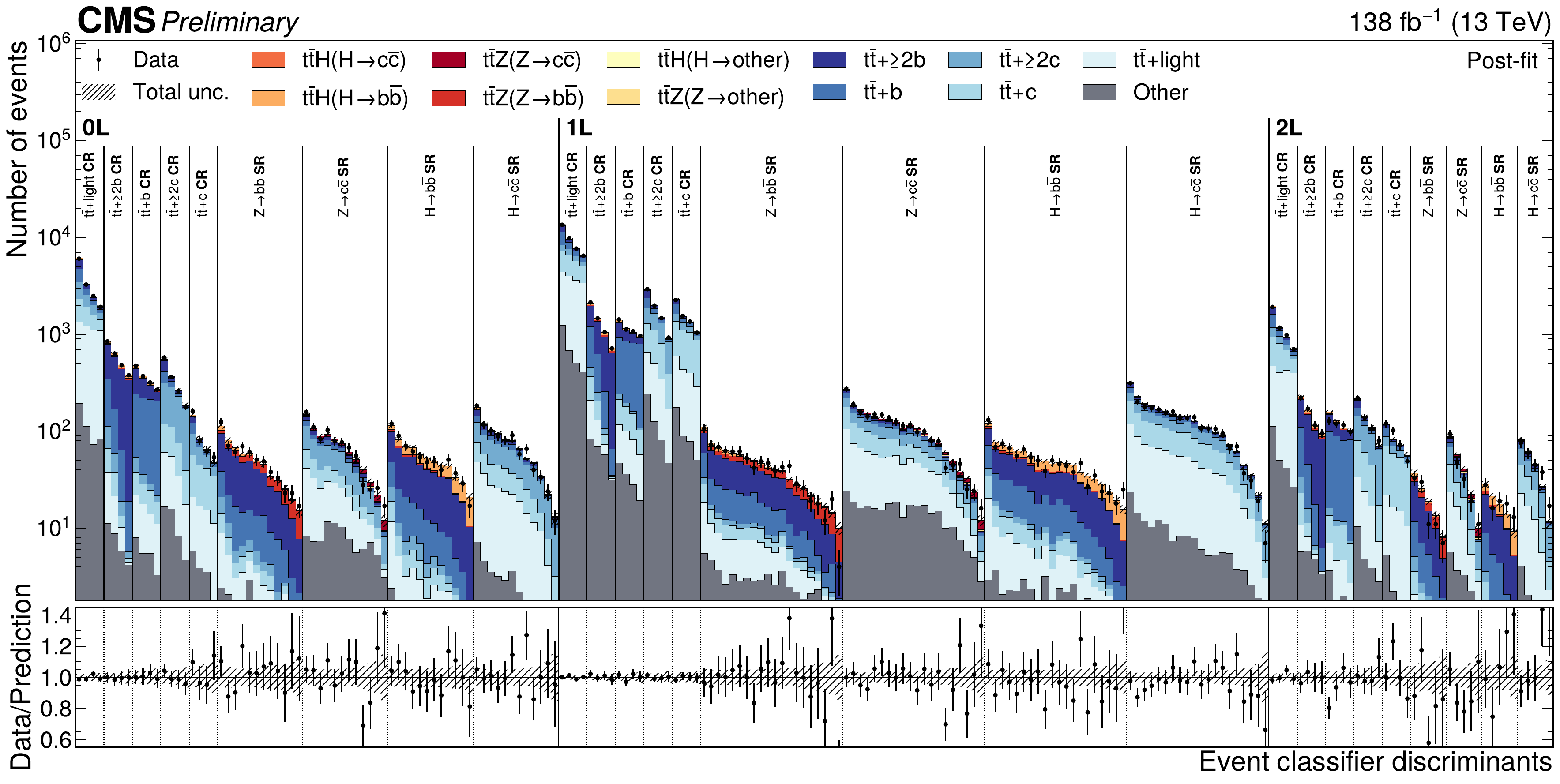}
\includegraphics[width=0.33\textwidth]{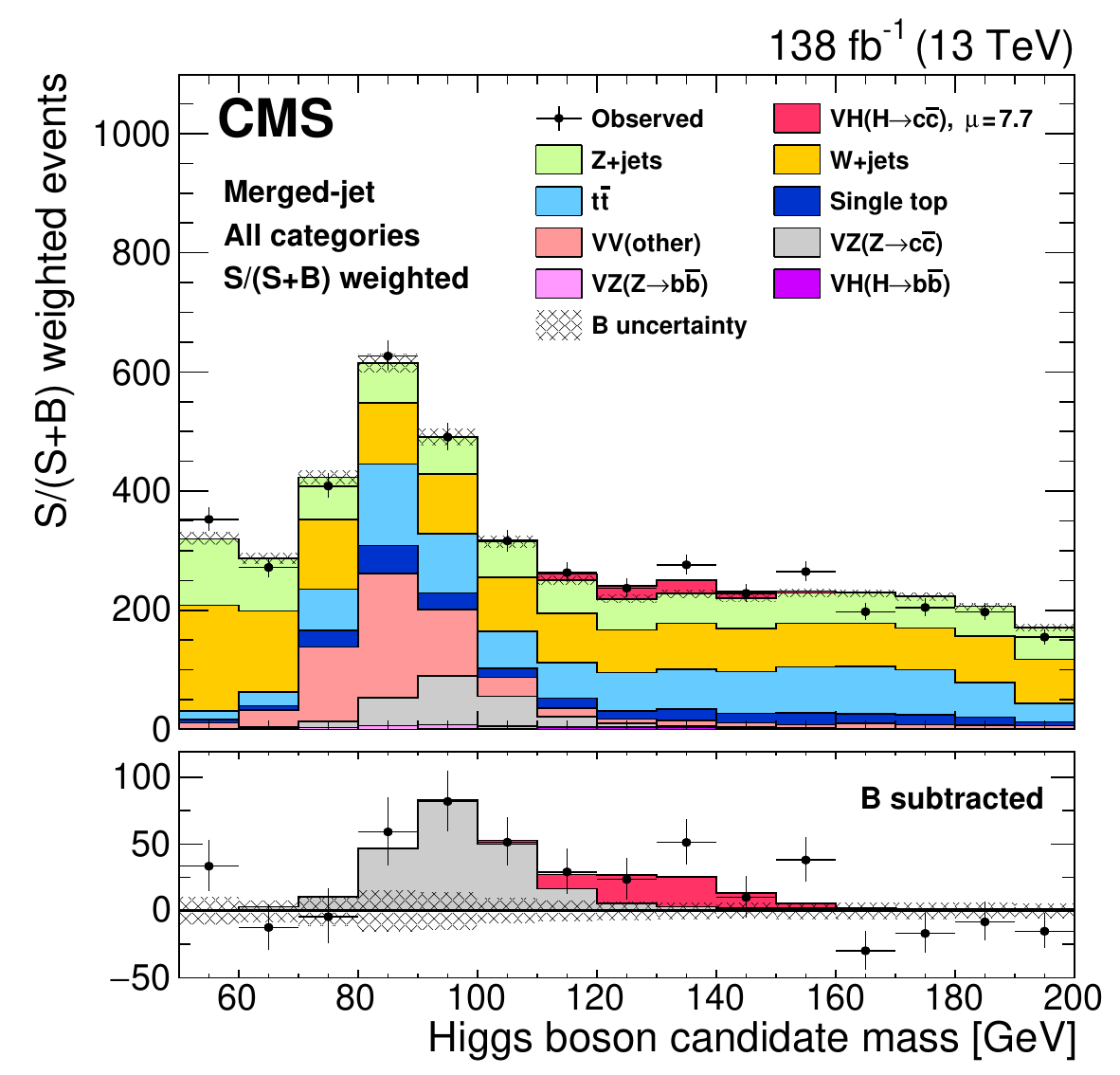}
\caption{Left: Post-fit discriminator outputs in each category and lepton multiplicity. The signals are represented by the four red and orange histograms~\cite{HIG24018}. Right: dijet invariant mass distribution in a representative $VH$ signal region~\cite{PRL131}.}
\label{fig:ttHcc}
\end{figure}

A simultaneous fit to the data is performed across the 9 categories, multiplied by the lepton multiplicity (0, 1, or 2 leptons). The regions enriched in the irreducible $t\bar{t}+b\bar{b}$, and $t\bar{t}+c\bar{c}$ backgrounds are included in the fit, and constrain the corresponding background processes directly from data. The post-fit distributions are shown in Figure~\ref{fig:ttHcc}. The sensitivity of the analysis is currently represented by the limited amount of data. The main systematic uncertainties affecting the result come from the modeling of the $t\bar{t} + c$ background.

The similar process $t\bar{t}Z$, with the $Z$ decaying to $b$ or $c$ quarks, is used to validate the fit strategy, and their measurements are consistent with the expectation from the SM.

The analysis is also very sensitive to the $t\bar{t}H, \, H\to b\bar{b}$ process, which can be measured with a signal strength $\mu = 0.91^{0.26}_{-0.22}$ and a statistical significance of 4.4 (observed) and 4.5 (expected) standard deviations.
The $t\bar{t}H, \, H\to c\bar{c}$ yield is too small to be measured, and an upper limit of 7.8 (observed) and 8.7 (expected) times the SM sit set at 95\% CL.

\subsection{Associated Production with a Vector Boson}

The $VH$ channel, where $V$ is a collective name for $W$ and $Z$ bosons, targets events where the Higgs boson is produced alongside a $W$ or $Z$ boson decaying leptonically, providing an efficient trigger and significant multijet background suppression. The CMS analysis~\cite{PRL131} employs the same \texttt{ParticleNet} tagger~\cite{ParticleNet2020} used in the $t\bar{t}H$ analysis to distinguish $c$-jets from $b$-jets and light-flavor jets. A ``boosted'' category is introduced, where the Higgs boson is produced with a sufficiently large Lorentz boost that the two $c$ quarks hadronize in a single, large-radius cone jet.
Events are divided into categories based on the number of identified leptons (0, 1, or 2) and the number of charm-tagged jets. A maximum-likelihood fit to the dijet invariant mass distribution is performed simultaneously across all categories. Figure~\ref{fig:ttHcc} shows an example of the invariant mass distribution in the signal region.

The analysis is able to observe, for the first time at a hadronic collider, the similar $VZ, \, Z\to c\bar{c}$ process with a statistical significance larger than 5 standard deviations and in excellent agreement with the SM: $\mu = 1.01^{0.23}_{-0.21}$. An upper limit on the more rare signal $VH, \, H\to c\bar{c}$ is set at 14 (observed) and 7.6 (expected) times the SM. This constrains the interval that the paramers $\kappa_c$ can assume to the range $1.1 < |\kappa_c| < 5.5$ at 95\% CL.

\subsection{Combination of direct charm Yukawa searches}

\begin{figure}[ht]
    \centering
    \includegraphics[width=0.6\textwidth]{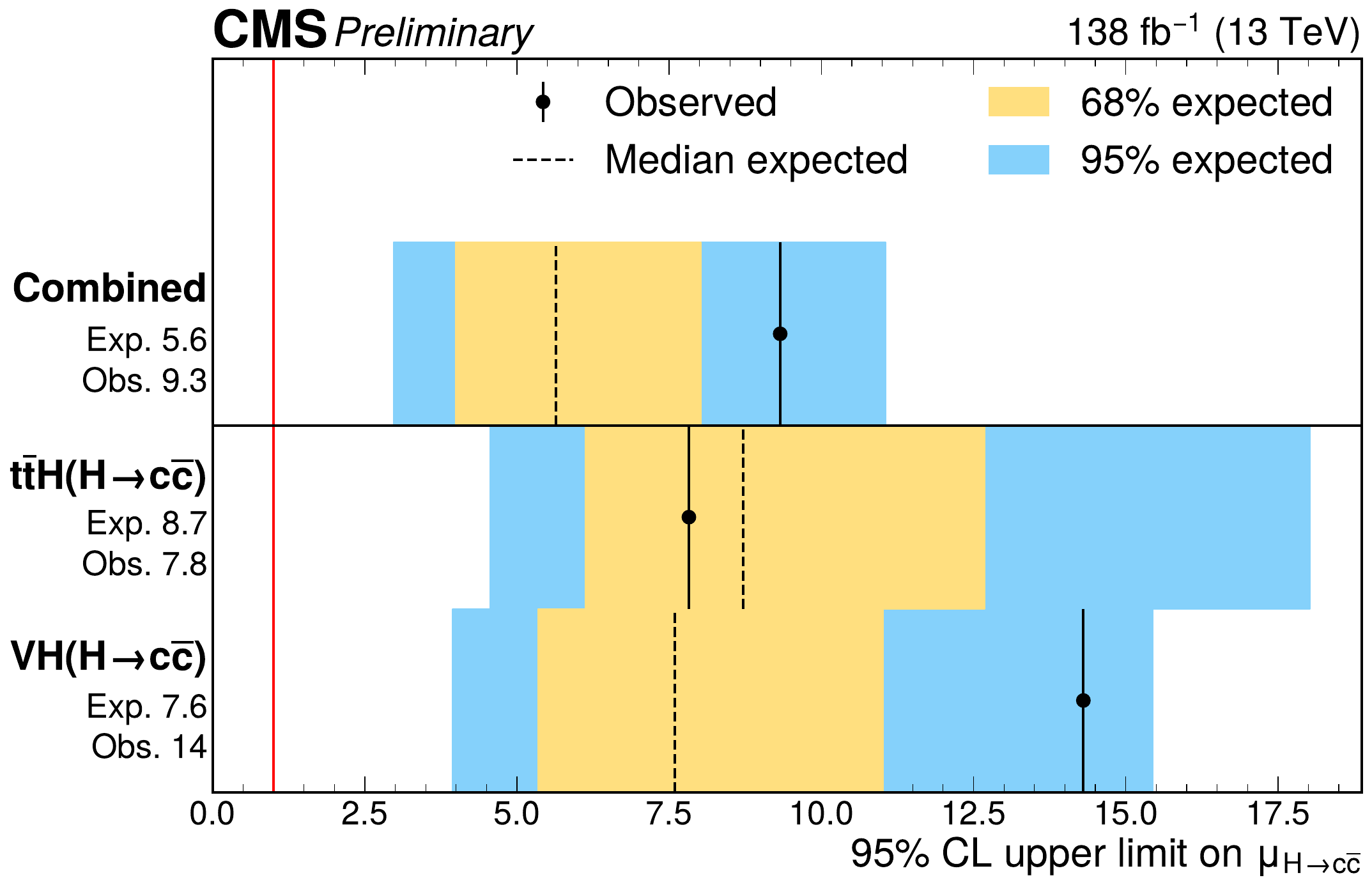}
    \includegraphics[width=0.35\textwidth]{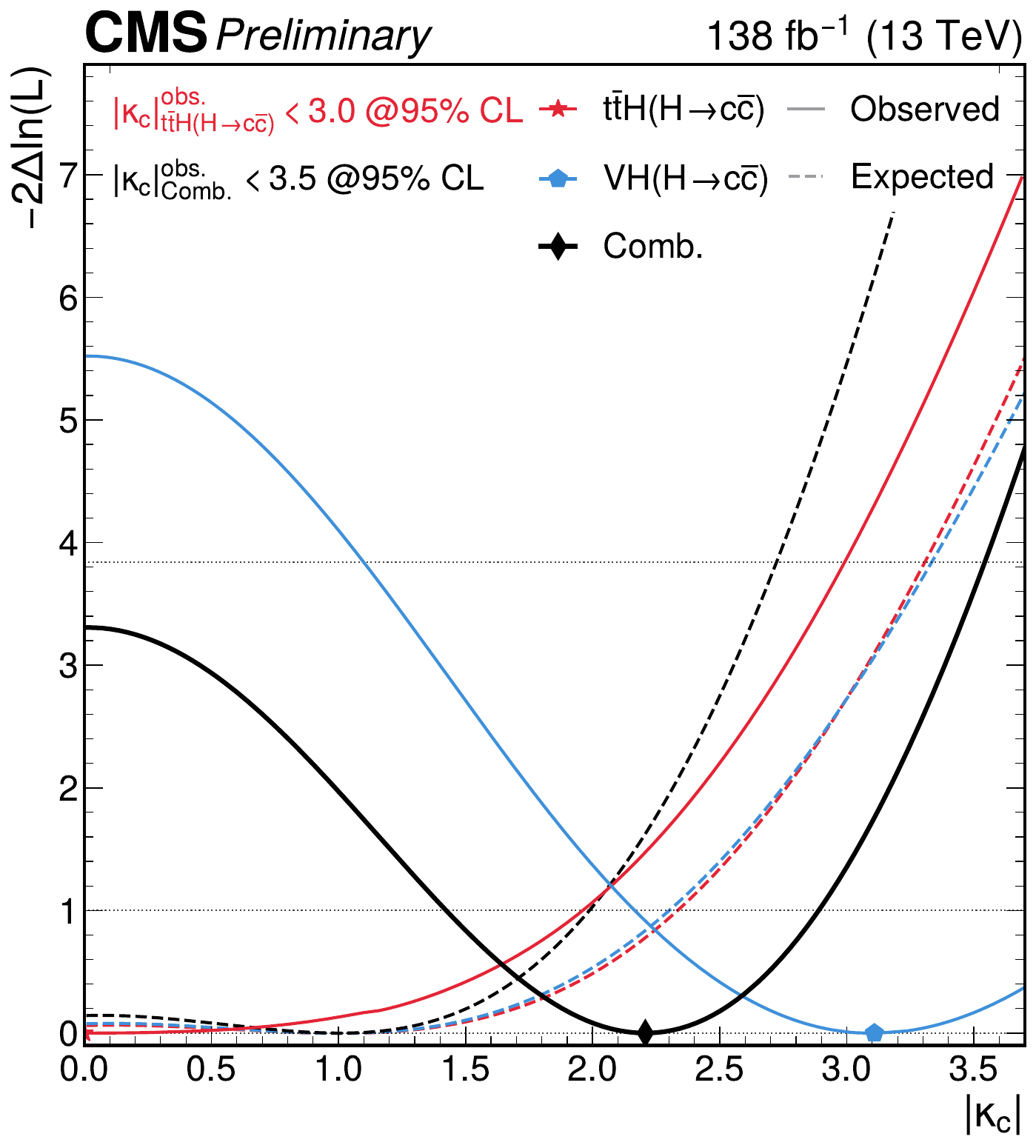}
    \caption{Observed and expected limits on $\sigma/\sigma_{SM}$ (left) and likelihood scan for the charm Yukawa coupling modifier $\kappa_c$ (right) for the $VH$ and $t\bar{t}H$, and the combination of the two~\cite{HIG24018}.}
    \label{fig:combi}
\end{figure}

The CMS Collaboration statistically combined the $VH$~\cite{PRL131} and $t\bar{t}H$~\cite{HIG24018} $H\to c\bar{c}$ analyses to provide the most stringent constraints on $\kappa_c$. In the statistical treatment, experimental and theoretical uncertainties are treated as correlated, while tagging uncertainties are considered as uncorrelated.
The profile of the likelihood scan of $|\kappa_c|$ and the representation of the combined upper limit on the combined signal strength is shown in Figure~\ref{fig:combi}.
With the combination, the allowed range for $|\kappa_c|$ is constrained to be below 3.5 (observed) and 2.7 (expected), representing the most stringent limit to date.

\section{Higgs Production in Association with a Charm Quark}
A complementary way to probe $y_c$ is through searches for the process $pp \to cH$, where the Higgs boson is produced in association with a single charm quark. In the SM, this process has a cross section of about $90$~fb, but new physics can significantly enhance the cross section and alter the Higgs boson $p_T$ spectrum, making the signal much simpler to distinguish from the SM background. CMS has searched for $cH$ production in both $H\to \gamma\gamma$~\cite{HIG23010} and $H\to WW^*$~\cite{HIG24009} final states.

\subsection{Diphoton channel}

In the $H\to \gamma\gamma$ channel~\cite{HIG23010}, events are selected with two isolated high-$p_T$ photons and at least one charm-tagged jet. The latter is identified using a $c$-tagger algorithm based on a Deep Neural Network, called \texttt{DeepJet}. Two Boosted Decision Trees (BDT) are trained to separate the signal from the two irreducible backgrounds, the continuous $\gamma\gamma$ production from QCD processes and the Higgs boson produced through gluon fusion, and define two categories. As shown in Figure~\ref{fig:cH}, the diphoton mass spectrum is fitted on both categories, yielding an upper limit on $|\kappa_c| < 38.1$ (observed) and 72.5 (expected).

\subsection{Diboson channel}

\begin{figure}[ht]
\centering
\includegraphics[trim=0 80 0 0,clip,width=0.36\textwidth]{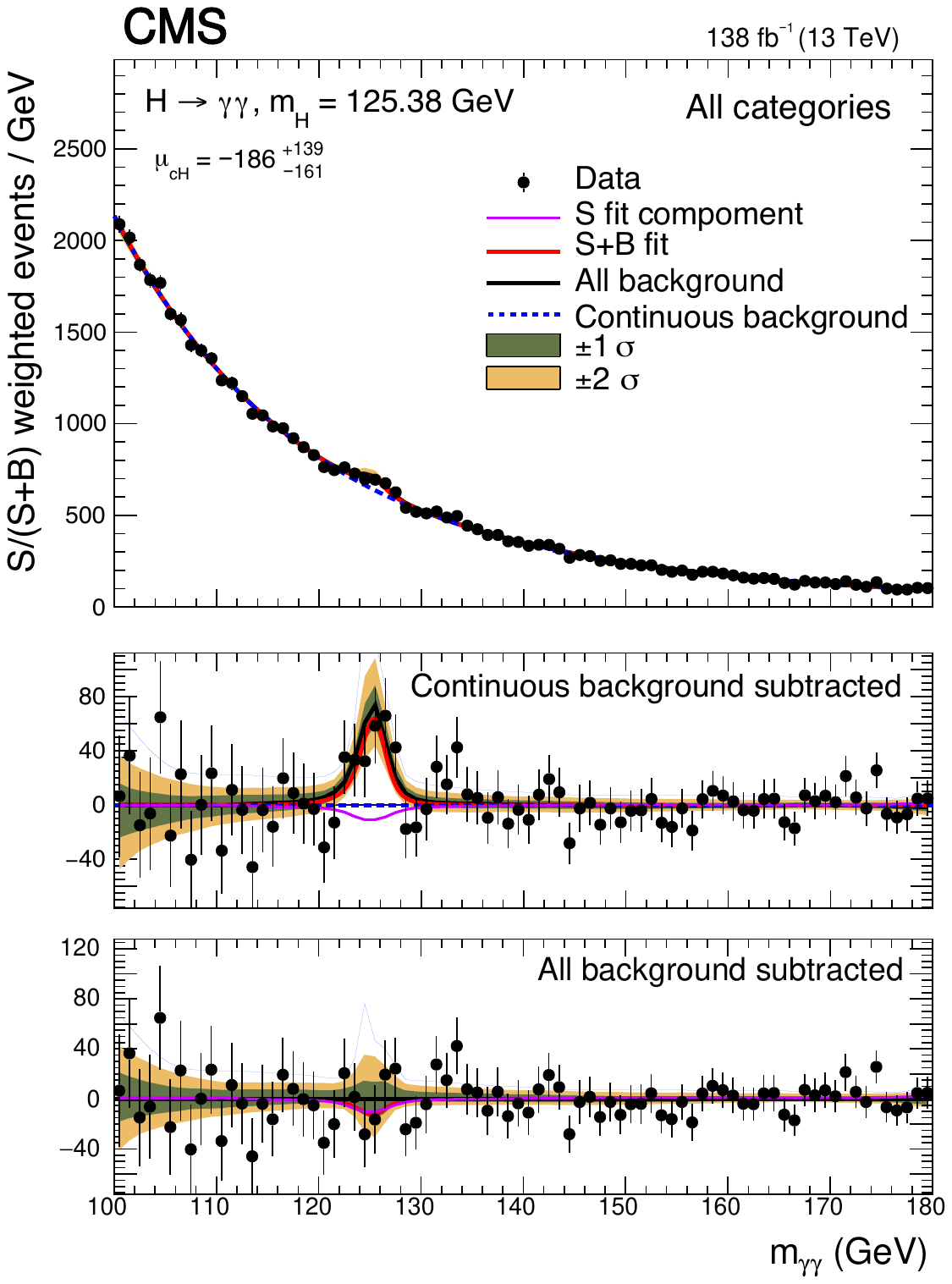}
\includegraphics[width=0.45\textwidth]{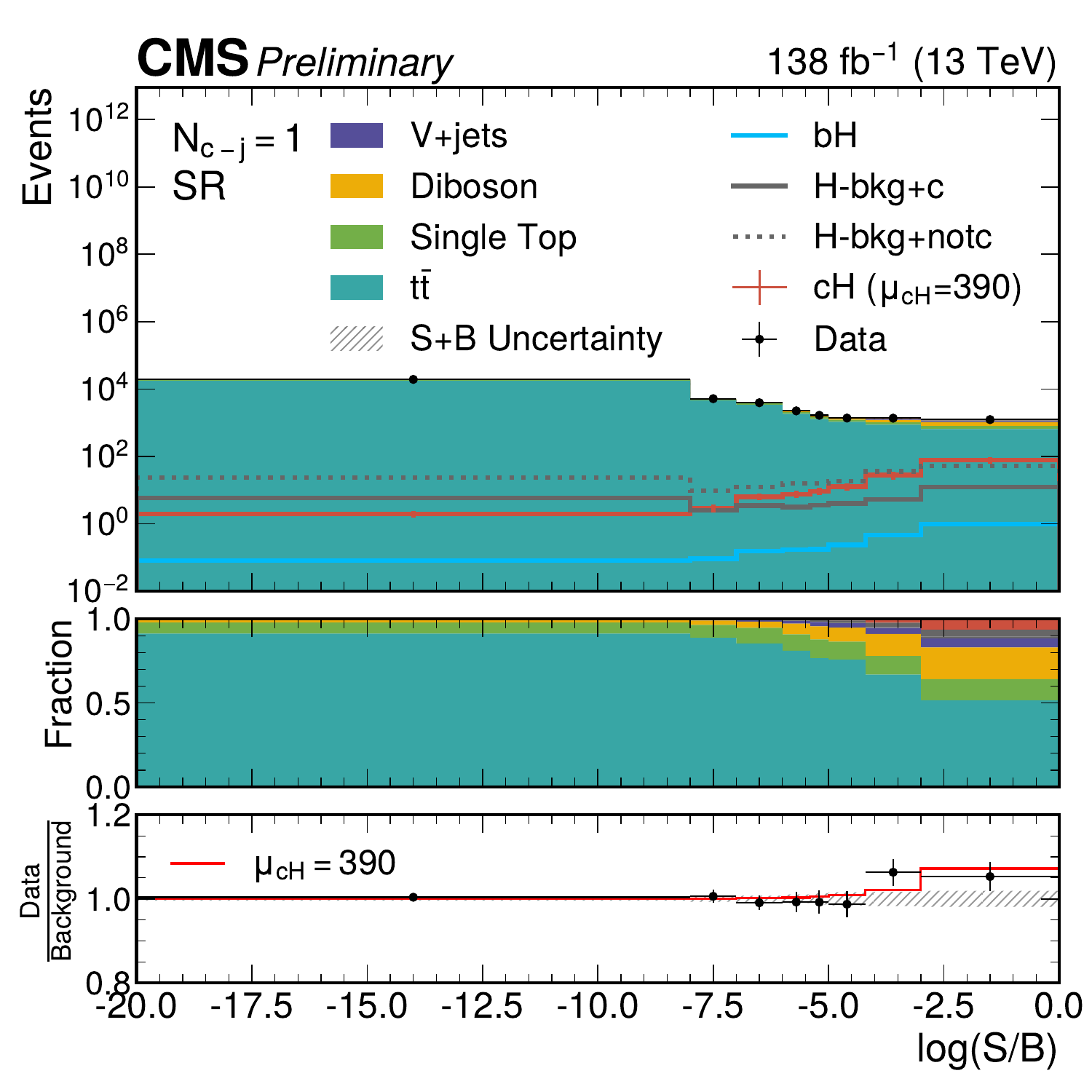}
\caption{Left: di-photon invariant mass spectrum in the $c H, \, H\to\gamma\gamma$ analysis~\cite{HIG23010}. Right: Weighted distribution of the events in the category with one charm-tagged jet in the $c H, \, H\to WW^*$ analysis~\cite{HIG24009}.}
\label{fig:cH}
\end{figure}

The search of the associated production with the Higgs boson decaying to a pair of $W$ bosons focuses on the $e\nu\mu\nu$ final state~\cite{HIG24009}, which is dominated by $t\bar{t}$ SM production. Also in this search, The \texttt{DeepJet} algorithm is used to select at least one $c$-tagged jet, and two BDTs are trained to optimized the signal-background separation (Fig.~\ref{fig:cH}). The constrains from the fit to the data are less stringent than the previous analysis, excluding $|\kappa_c| < 211$ (observed) and 95 (expected).

\section{Indirect Probes via Radiative Decays}

\subsection{Charm loop mediated decays}

\begin{figure}[ht]
    \centering
    \includegraphics[width=0.46\textwidth]{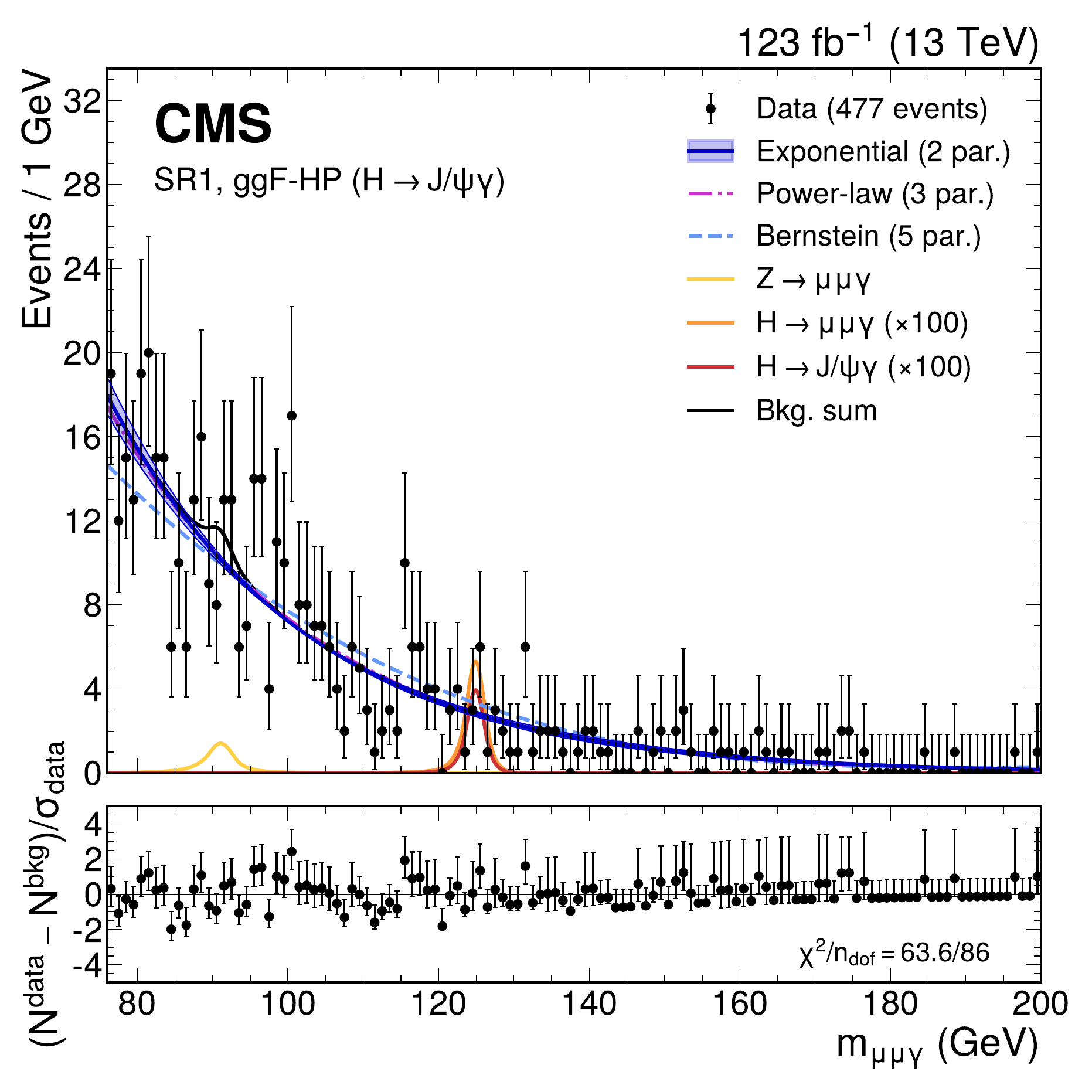}
    \includegraphics[width=0.46\textwidth]{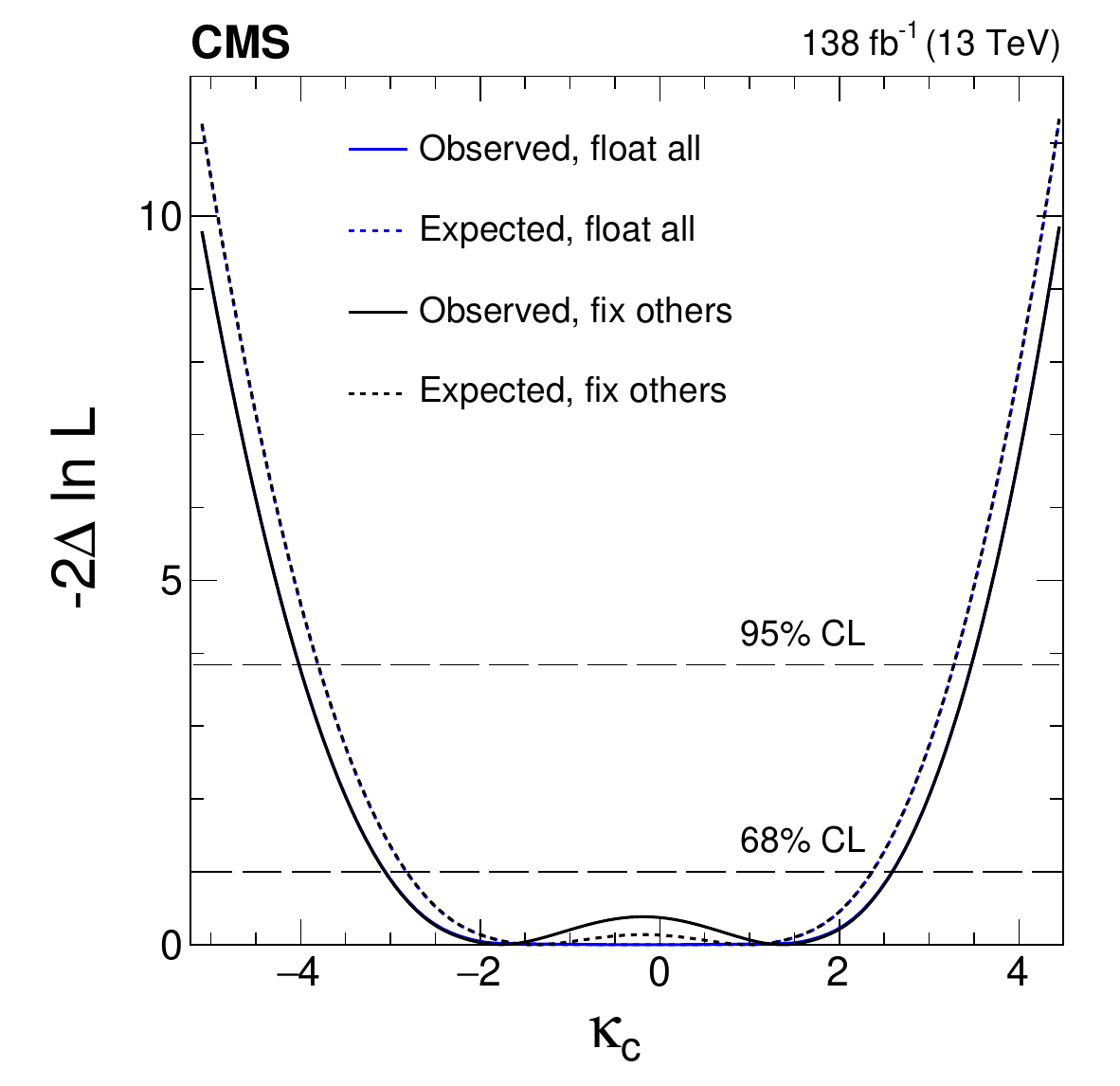}
    \caption{Left: invariant mass distributions in the search for $H\to J/\Psi \gamma$~\cite{PLB865}. Right: Likelihood scan on $\kappa_c$~\cite{HIG23011}. The black lines represent the case where all couplings except $\kappa_c$ are fixed at their SM values.}
    \label{fig:meson}
\end{figure}

The decay of the Higgs boson to a $J/\Psi$ or a $\Psi'$ meson (collectively referred to as $\Psi(nS)$) and a photon represents a probe to the charm Yukawa coupling, because the decay may occur through a loop of charm quarks~\cite{PLB865}. Direct and non-direct diagrams interfere destructively, the latter being approximately 18 times larger in magnitude than the former. The final state consists of a photon and two muons from the $\Psi(nS)$ decay, a signature that is difficult to mimic by the copious multijet background. This analysis exploits the spin correlations in the signal to perform an angular analysis of the decay, and uses a categorization targeting different Higgs production modes, which further increases sensitivity to VBF and Higgs associated production processes.
A fit to the three-body invariant mass spectrum with parametric functions (Figure~\ref{fig:meson}) is used to set  limits on the branching fraction of these decays: $B(H\to J/\Psi\gamma) < 2.6\times 10^{-4}$ and $B(H\to \Psi'\gamma) < 9.9\times 10^{-4}$ at 95\% CL. By assuming that all other couplings are fixed at the SM value, the bound on the coupling is found to be $-166 < \kappa_c < 208$. An upper limit of approximately 7 times the SM is placed on the  similar $Z \to\ J/\Psi\gamma$ decay.

\subsection{Indirect constraints from $\gamma H$}

Constraints on the coupling of the Higgs boson to the charm quark can also be placed by accurately measuring other known processes involving the Higgs boson, as alterations of the couplings would enhance their production rates. By considering the well known $H\to ZZ^*\to 4\ell$, both from gluon fusion and the associated $\gamma H$ production, a CMS analysis~\cite{HIG23011} places stringent constraints on $-4.0 < \kappa_c < 3.5$, assuming all other couplings are as predicted by the SM.

\section{Summary}

Exploring the charm-Higgs coupling is the next challenge for the LHC experiments, representing a crucial milestone towards a complete understanding of the Higgs mechanism across all fermion generations. With Run~2 data, a complete set of results has been published by CMS searching for the yet-to-be-observed charm-Higgs coupling; these results are consistent with the SM expectations. Run~3 of the LHC and the HL-LHC program will provide unprecedented datasets, providing the statistical precision needed to observe rare Higgs processes and shed light on this elusive coupling.

An equally important aspect of this effort is the rapid progress in machine-learning-based jet flavor identification for charm-jet tagging, which has led to substantial performance improvements in recent years, far beyond the increase in the integrated luminosity. The introduction of state-of-the-art neural network architectures, such as transformer networks employed in the $t\bar{t}H,\ H\to c\bar{c}$ analysis, has enabled the classification of complex event topologies that were previously inaccessible, thereby opening new channels for these searches.

\end{document}